\documentclass[preprint,12pt]{elsarticle}

\usepackage{graphicx}
\usepackage{subfigure}
\usepackage{amssymb}

\usepackage{mathtools}

\journal{Journal Name}

\begin{document}

\begin{frontmatter}


\title{A lateral semicircular canal segmentation based geometric calibration for human temporal bone CT Image}



\author{Xiaoguang Li\textsuperscript{a,c,*}}
\author{Peng Fu\textsuperscript{a}}
\author{Hongxia Yin\textsuperscript{b*}}
\author{Zhenchang Wang\textsuperscript{b}}
\author{Li Zhuo\textsuperscript{a,c}}
\author{Hui Zhang\textsuperscript{a,c}}

\address{\textsuperscript{a}Faculty of Information Technology, Beijing University of Technology, Beijing 100124, China}
\address{\textsuperscript{b}Department of Radiology, Beijing Friendship Hospital, Capital Medical University, Beijing, 100050, China}
\address{\textsuperscript{c}Beijing Key Laboratory of Computational Intelligence and Intelligent System, Beijing University of Technology, Beijing 100124, China}

\begin{abstract}
Computed Tomography (CT) of the temporal bone has become an important method for diagnosing ear diseases. Due to the different posture of the subject and the settings of CT scanners, the CT image of the human temporal bone should be geometrically calibrated to ensure the symmetry of the bilateral anatomical structure. Manual calibration is a time-consuming task for radiologists and an important pre-processing step for further computer-aided CT analysis. We propose an automatic calibration algorithm for temporal bone CT images. The lateral semicircular canals (LSCs) are segmented as anchors at first. Then, we define a standard 3D coordinate system. The key step is the LSC segmentation. We design a novel 3D LSC segmentation 
encoder-decoder network, which introduces a 3D dilated convolution and a multi-pooling scheme for feature fusion in the encoding stage. The experimental results show that our LSC segmentation network achieved a higher segmentation accuracy. Our proposed method can help to perform calibration of temporal bone CT images efficiently.
\end{abstract}


\end{frontmatter}


\section{Introduction}
\label{S:1}

Temporal bone Computed Tomography(CT) imaging is an established standard for the examination of ear diseases \cite{jager2005ct}. Due to the differences in patient postures, the raw CT images of the temporal bone are not aligned. Radiologists have to manually reconstruct the temporal bone CT slices so that the bilateral anatomies are symmetrical. After that, the CT slices will be sent to doctors to diagnose. With the increasing of temporal bone CT images, to radiologists, manually calibration becomes one of a time-consuming burden. However, temporal bone CT calibration is an important pre-processing step for further computer-aided CT analysis. In addition, uncalibrated CT images of the raw temporal bone pose challenges to the establishment of a standardized temporal bone CT database, which is not conducive to medical research. Therefore, develop an automatic temporal bone CT calibration method is of great significance to clinical practice and medical research.\par
\begin{figure}
\centering
\includegraphics[scale=0.75]{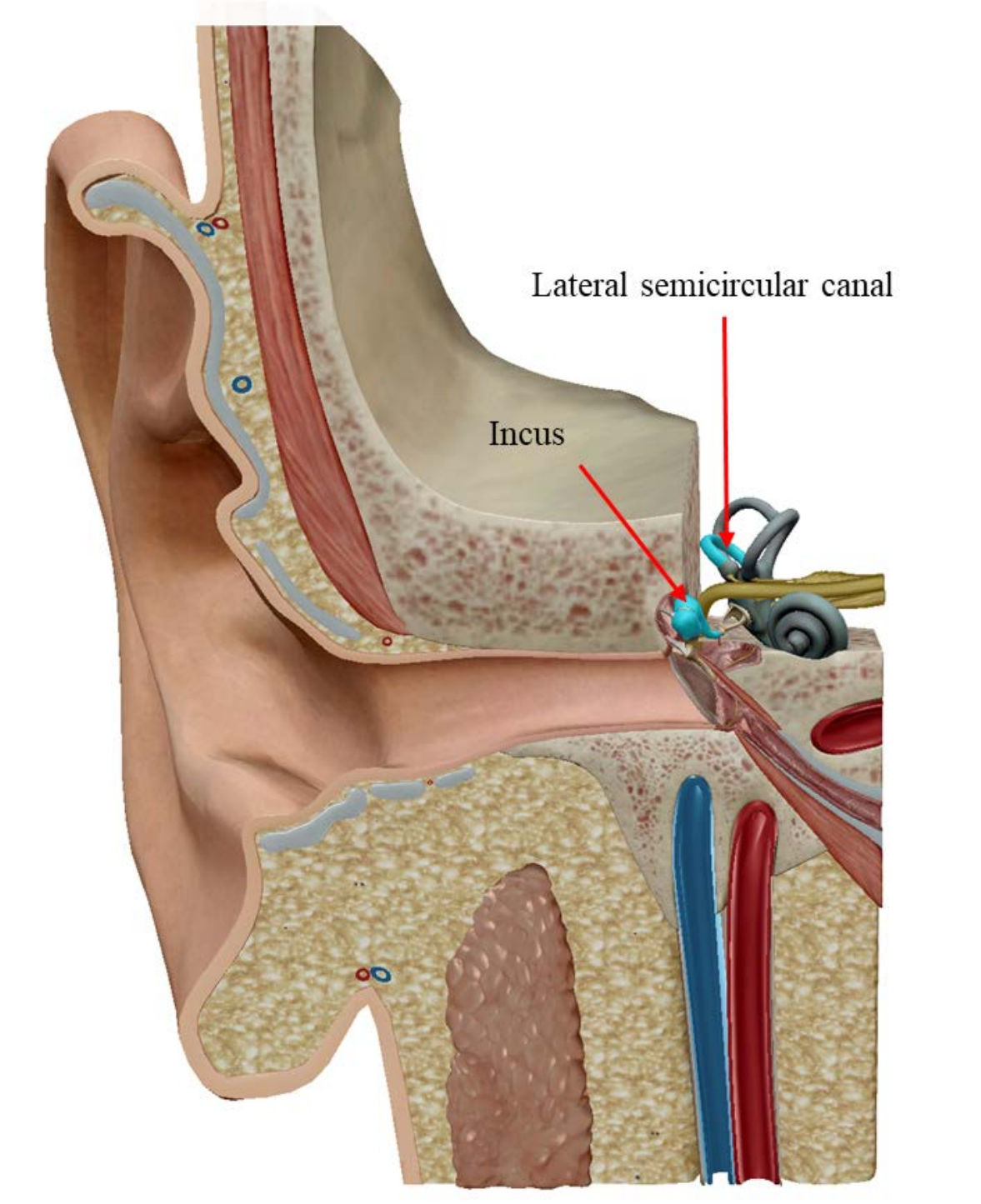}
\caption{Anatomical illustration of critical anatomical organs in the temporal bone,
they are bilaterally symmetrical in the human ear.}
\label{Fig:1}
\end{figure}

The anatomical structure of the human temporal bone is shown in Figure. \ref{Fig:1}, where the key anatomical organs are approximately bilaterally symmetrical. To present bilateral asymmetry view of the raw temporal bone CT images, radiologists should adjust the reconstructed view carefully so that the two lateral semicircular canals (LCSs) can be shown in the same axial slice symmetrically. The calibration steps are as follows. Firstly, find out the slice plane that shows the full semicircular shape of the LSCs; Then, reconstruct the lateral semicircular canal parallel to the sagittal plane on the cross-sectional reconstruction line, and fine-tune the axial image to the structural symmetry on both sides; Finally choosing the double-sided incus joint conforms to the best aspect of the body display.\par

Inspired by manual calibration procedure of the raw temporal bone CT images, we find that the lateral semicircular canals are important anchors. We select the left and right lateral semicircular canals as the anchors to guide the automatic calibration to eliminate the asymmetry of the bilateral structure.\par
\begin{figure}[htbp]
\centering
\subfigure[Transverse position]{
\begin{minipage}[t]{0.85\linewidth}
\centering
\includegraphics[scale=.65]{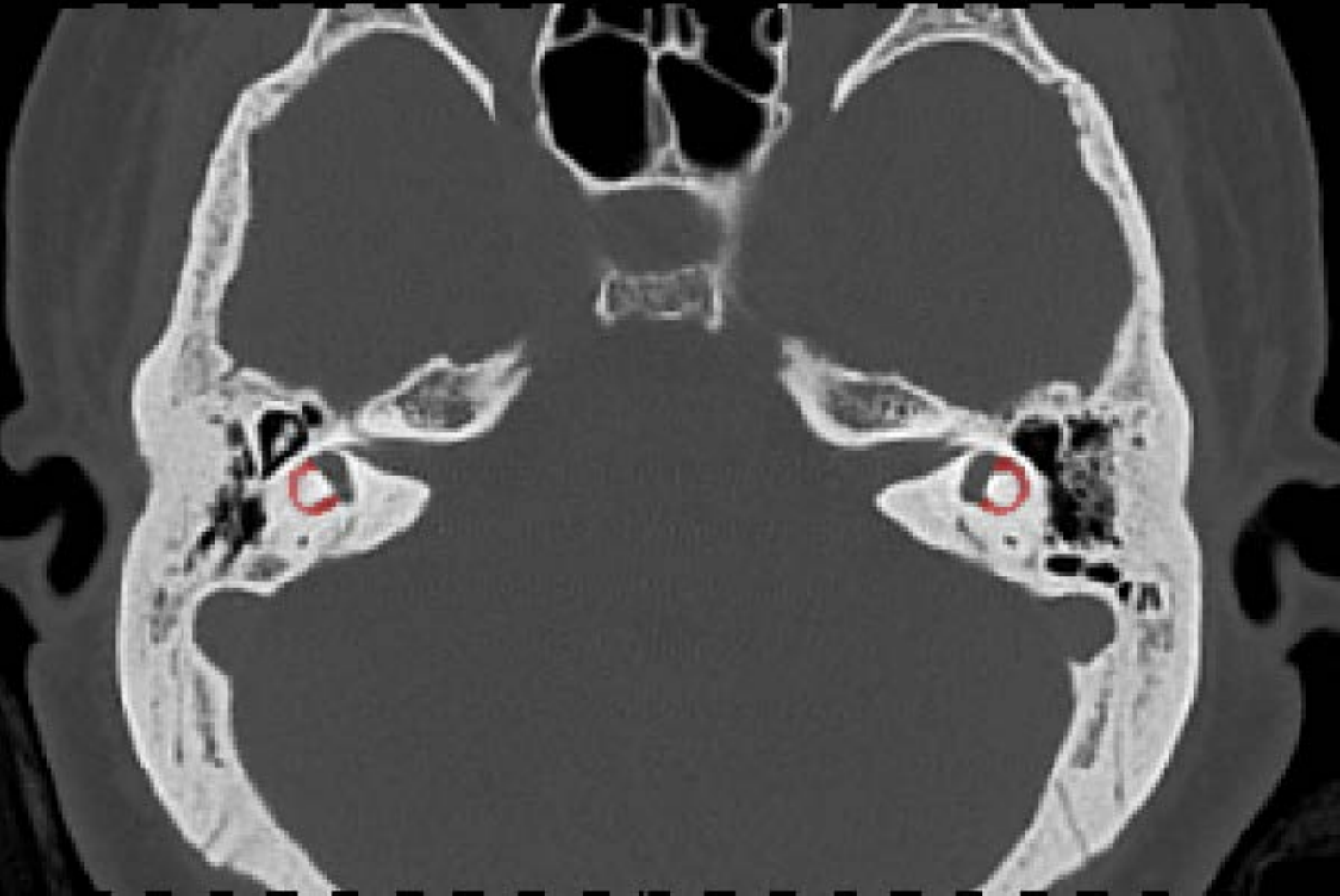}
\end{minipage}%
}%

\subfigure[Coronal position]{
\begin{minipage}[t]{0.75\linewidth}
\centering
\includegraphics[scale=.55]{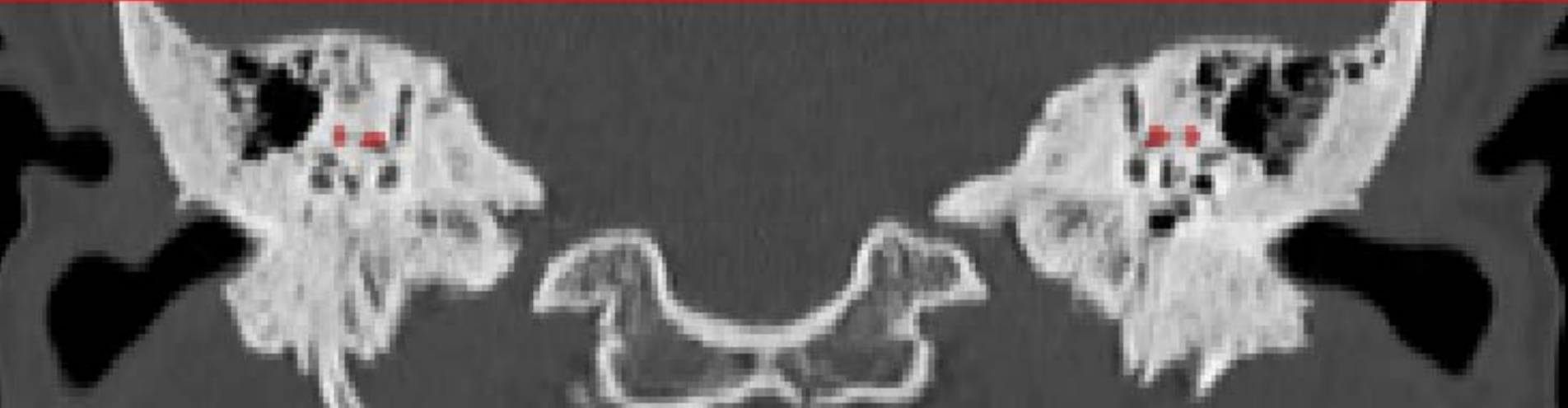}
\end{minipage}%
}%

\subfigure[Sagittal position]{
\begin{minipage}[t]{1\linewidth}
\centering
\includegraphics[scale=.65]{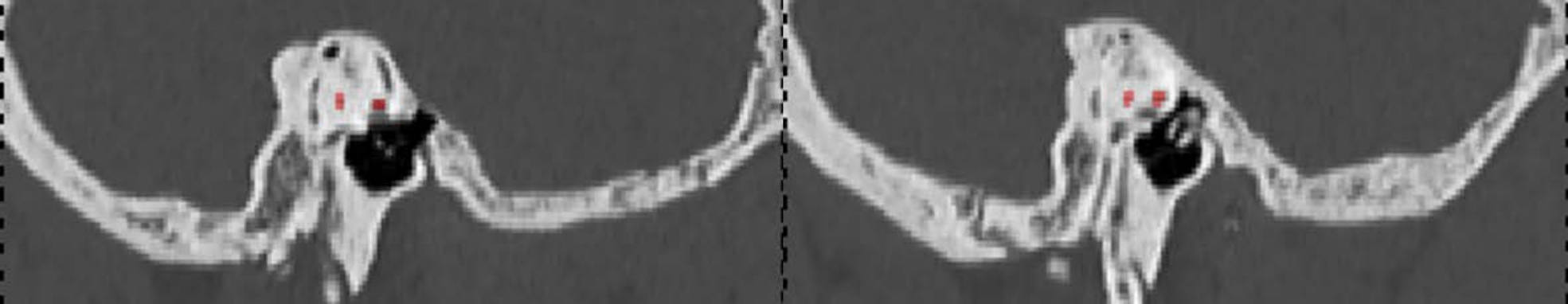}
\end{minipage}
}%

\centering
\caption{The LSCs in different views.}
\label{Fig:2}
\end{figure}

Annotation of the lateral semicircular canal plays an important role in the automatic calibration method of the raw temporal bone CT image. However, large-scale medical image data annotation is expensive and difficult to obtain. Therefore, we proposed an LSC segmentation algorithm to roughly predict the location of anchors. The lateral semicircular canal is a small sophisticated anatomical structure in the temporal bone. It is a semicircle canal that embedded in the temporal bone. In a CT volume, a lateral semicircular canal contains about 200 to 300 voxels. In different observation positions, the LSC vary greatly, as shown in Figure \ref{Fig:2}, which also makes it difficult to segment accurately.\par

We propose a lateral semicircular canal segmentation based automatic geometric calibration method for human temporal bone CT images. Firstly, a novel LSC segmentation, 3D Multi Feature Fusion Network (3D-MFF Net) is proposed. This algorithm automatically segments the LSC in the raw temporal bone CT images; Then, we conduct a geometric calibration and resampling to automatic calibration.\par
The contributions of this work are summarized as follows:
\begin{enumerate}[(1)]
\item A novel automatic geometric calibration method for the raw temporal bone CT images is proposed for the first time. Inspired by the manual calibration steps, we select the lateral semicircular canals as anchors to estimate the parameters for affine transformation. The calibration method can adjust the reconstruction view to ensure the left and right LSCs are shown in the same slice in axial position.
\item As the key step of the calibration method, we propose a novel LSC segmentation algorithm, 3D-MFF net, for the segmentation of the LSC robustly. In this algorithm, we designed a multi-scale and multi-mode pooling feature fusion module to reduce the loss of information caused by pooling operations.
\item A dilated convolution module is introduced into the 3D network to extract spatial semantic information at different scales, to improve the accuracy of the LSC segmentation.
\item We supply extensive performance comparisons. Experimental results and analysis show that our method can achieve satisfied calibration results for raw temporal bone CT images. It has great potential to reduce the workload of radiologists and provides an important foundation for raw temporal bone CT images pre-processing.
\end{enumerate}

\section{Related work}

\subsection{CT images calibration}

The raw temporal bone CT image is affected by the patient's posture. The imaging shows different degrees of skew, and the bilateral structures in the temporal bone are asymmetric. Therefore, the raw temporal bone CT image needs to be reconstructed in multiple planes, so that the bilateral structure in the reconstructed temporal bone CT is symmetrical.\par

In medical research, to measure the direction of the human semicircular canals \cite{della2005orientation}, the researchers manually reconstruct the temporal bone CT of 22 human subjects through 3D rotation to achieve multi-plane reconstruction, and finally make the image plane in the CT roughly coplanar with the semicircular canal. Manual calibration limits the sample size of the study to small examples. In clinical practice, imaging physician needs to perform a multi-plane reconstruction of the raw temporal bone CT data on a post-processing workstation. This process also needs to be done manually.\par

The multi-planar reconstruction of temporal bone CT is essentially a 3D rotation of the CT data. This step can be successfully completed by manual operation, but this greatly increases the workload, therefore, there are urgent demands for computer-aided technology to automatically calibrate the raw temporal bone CT. However, as far as we know, we have not found work in this topic. Automatic calibration technology is different from manual calibration. Manual calibration can subjectively fine-tune the bilateral structure in the temporal bone. The key step of our automatic calibration is the accurate segmentation of the LSCs.

\subsection{Medical images segmentation}

Segmentation is an important step in medical image analysis. It is dedicated to identifying lesion or organ voxels from the background, which usually plays an important role in the lesion assessment and disease diagnosis. At present, medical image segmentation methods mainly include traditional segmentation methods and deep-learning-based segmentation methods. \par

Traditional segmentation methods mainly include threshold methods, region growing methods and cluster-based methods. The threshold methods \cite{xie2010application} classify the pixels of an image by setting one or several specific thresholds. Among them, pixels within the same intensity threshold range belong to the same object. The selection of the threshold requires constant adjustment through experience and experiments. The region growing methods \cite{pohle2001segmentation} extract the connected regions in an image around the preset seed points. The selection of seed points is a challenge, which affects the accuracy of image segmentation significantly. As for the clustering methods, the segmentation accuracy mainly depends on the setting of the initial number of categories.\par

With the rapid development of deep learning technology, deep-learning-based segmentation methods have made a great breakthrough in medical image segmentation. A great number of deep neural network architectures are proposed for different specific segmentation tasks. The deep-learning-based segmentation methods have been widely used in segmentation tasks such as skin \cite{esteva2017dermatologist}, brain \cite{havaei2017brain}, heart \cite{chen2020deep}, lung, tumor, and so on \cite{wang2019abdominal}. Different architectures have been proposed. For example, some solve the gradient disappearance and explosion issues, some are designed to extract more abundant semantic information, or compress the model for resource limited applications.\par

Fully Convolutional Network (FCN) \cite{long2015fully} is a pioneering work of semantic segmentation. The FCN employed convolutional layers to replace the fully connected layers in a convolutional neural network to obtain output results with the same resolution as the input image and achieved pixel-level classification.
To improve the accuracy of the FCN, an encoder and decoder architecture, U-net, is popular in medical image segmentation \cite{ronneberger2015u}. In the U-net, the input image is down-sampled in the encoding stage, and up-sampled in the decoding stage until they are consistent with the input image resolution. Skip connections are also used between the encoding and decoding stages to recover detailed information and improve the accuracy of segmentation.\par

In medical image segmentation, U-net is one of the most successful network frameworks. Many U-Net variants are proposed by introducing novel modules or improvements. For example, Res-UNet \cite{xiao2018weighted} and Dense U-Net \cite{guan2019fully} are inspired by residual connections \cite{he2016deep} and dense connections \cite{huang2017densely}, and replace the modules in the encoding stage of the U-Net with residual connections and dense connection respectively. In neural networks, cascading convolution layers and pooling may reduce the spatial resolution of the input feature map. To address this issue, Gu \textit{et al}. \cite{gu2019net} proposed a context encoding network to extract richer semantic information. This context extraction module consists of a dense dilated convolution module and a residual multi-kernel pooling module, which extract richer context information through multi-scale dilated convolution and multi-scale pooling, respectively.\par

Different from natural images, most medical images are 3D data. To make full use of the 3D volume, 3D network architectures are designed. 3D-Unet \cite{cciccek20163d} extends operations in the U-net to corresponding 3D version. V-net \cite{milletari2016v} is another 3D version of U-net, which adds residual connection in the network and replaces pooling operation with convolution. To solve the problem of small organ segmentation, 3D-DSD \cite{li20203d}, a 3D dense connection blocks and 3D multi pooling feature fusion schemes are designed in encoding stage of the 3D-Unet, and adopted a 3D deep supervision mechanism in the decoding stage \cite{lee2015deeply}, to improve the accuracy of small organs.\par

Aim to the LSC segmentation task in this paper, we proposed a novel 3D network. Take the 3D-DSD as the baseline framework, we propose a 3D multi-scale dilated convolution and a multi-scale and multi-mode pooling feature fusion strategy in the encoding stage.

\section{The proposed method}

\begin{figure}
	\centering
		\includegraphics[scale=.65]{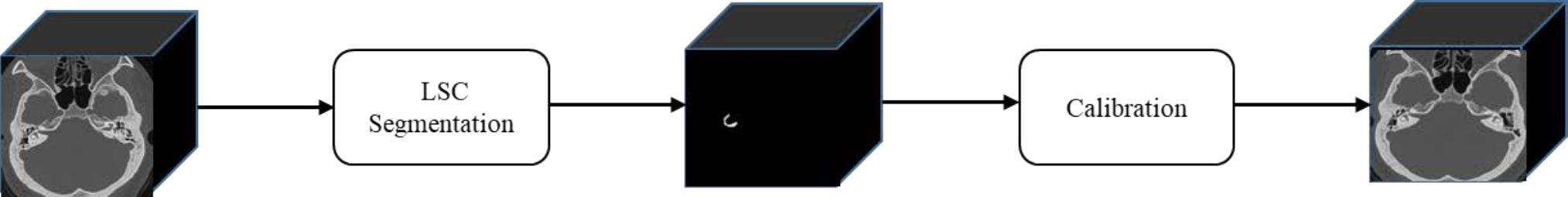}
	\caption{Overall architecture of the automatic calibration algorithm, which consists of a segmentation algorithm and a calibration algorithm.}
	\label{Fig:3}
\end{figure}

Similar to the eyes in a face image, the LSCs can be used as anchors in the temporal bone CT. In this section, we demonstrate the proposed automatic geometric calibration method for temporal bone CT images in detail. The framework of the calibration algorithm is shown in Figure \ref{Fig:3}. It includes two parts, the LSC segmentation and the geometric calibration. Firstly, the segmentation algorithm is used to automatically segment the lateral semicircular canal, and then the automatic calibration algorithm is used to estimate the parameter matrix for affine transformation. The calibration is implemented by translation, rotation and resampling on the raw temporal bone CT slices.\par

\subsection{Segmentation}
\begin{figure}
	\centering
		\includegraphics[scale=.55]{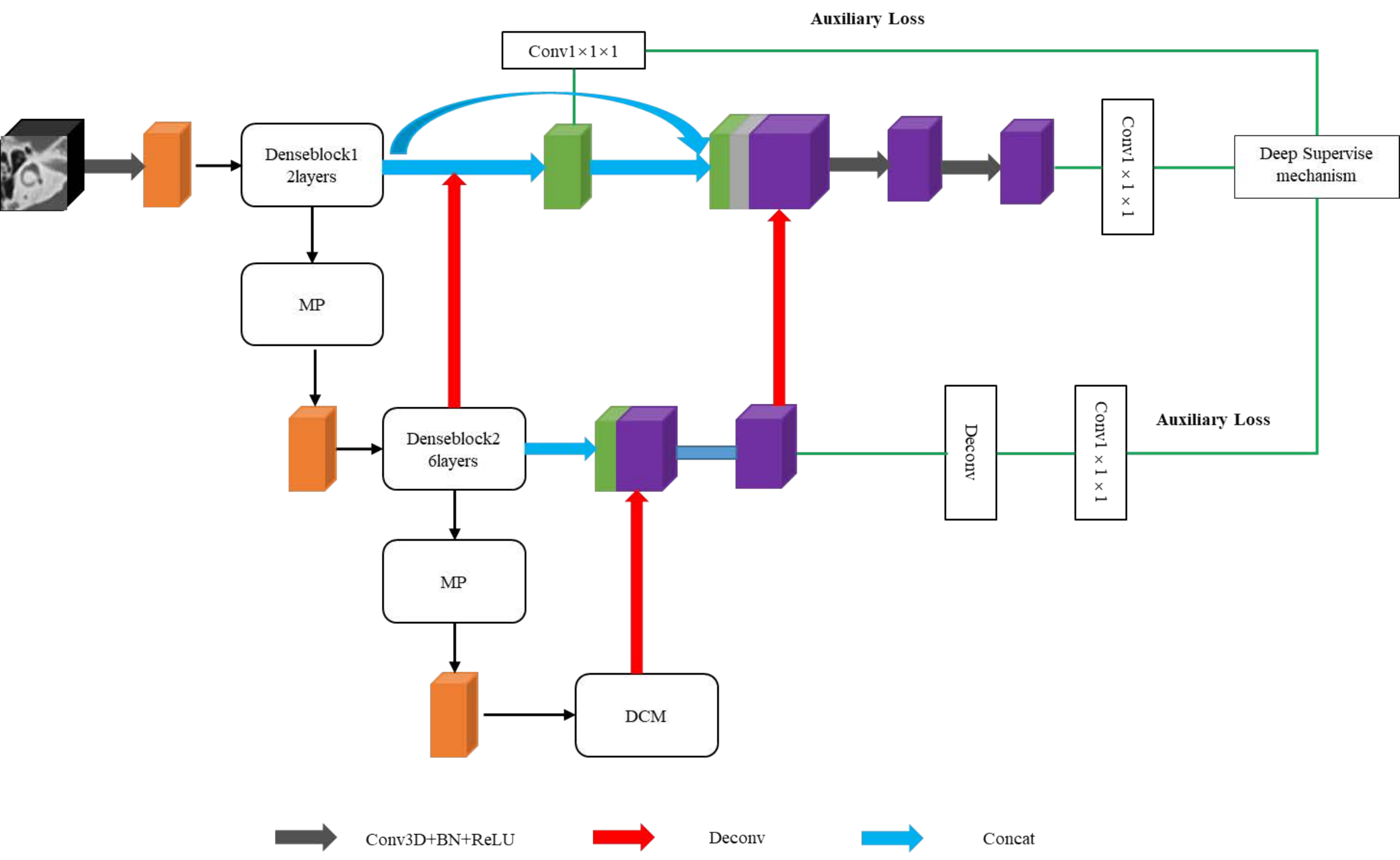}
	\caption{The proposed 3D-MFF is an encoder-decoder architecture. It employed two dense connection blocks and a dilated convolution module (DCM) in the encoder and a multi scale-mode pooling module (MP). A deep supervision mechanism is employed to train the network.}
	\label{Fig:4}
\end{figure}

Based on the 3D-DSD[18], we propose a novel network architecture, 3D-MFF, for LSCs segmentation. The network architecture is a U-net like network, as shown in Figure \ref{Fig:4}. Firstly, we introduced a 3D dilated convolution module to extract multi-scale high-level semantic information in the encoding stage. Secondly, a multi pooling fusion module is proposed to make full use of sematic information via different scale and pooling mode. Finally, employe the 3D deep supervision proposed by 3D-DSD to help training.

\subsubsection{Encoder}

\paragraph{(1)} Dilated convolution module\par

Segmentation for small targets is always a challenge for medical image segmentation. To improve the accuracy of the small target segmentation, it is very important for neural networks to make full use of the extracted multi-layer and multi-scale feature maps.\par

Inspired by Atrous Spatial Pyramid Pooling \cite{chen2017rethinking} and multi-scale dilated convolution in CE-Net, we introduce a 3D dilated convolution module to the encoder. The idea of the 3D dilated convolution module is to extract multi-scale information through dilated convolution with different expansion rates, and obtain richer semantic information in the encoding stage to improve the segmentation accuracy.\par

\begin{figure}
	\centering
		\includegraphics[scale=.55]{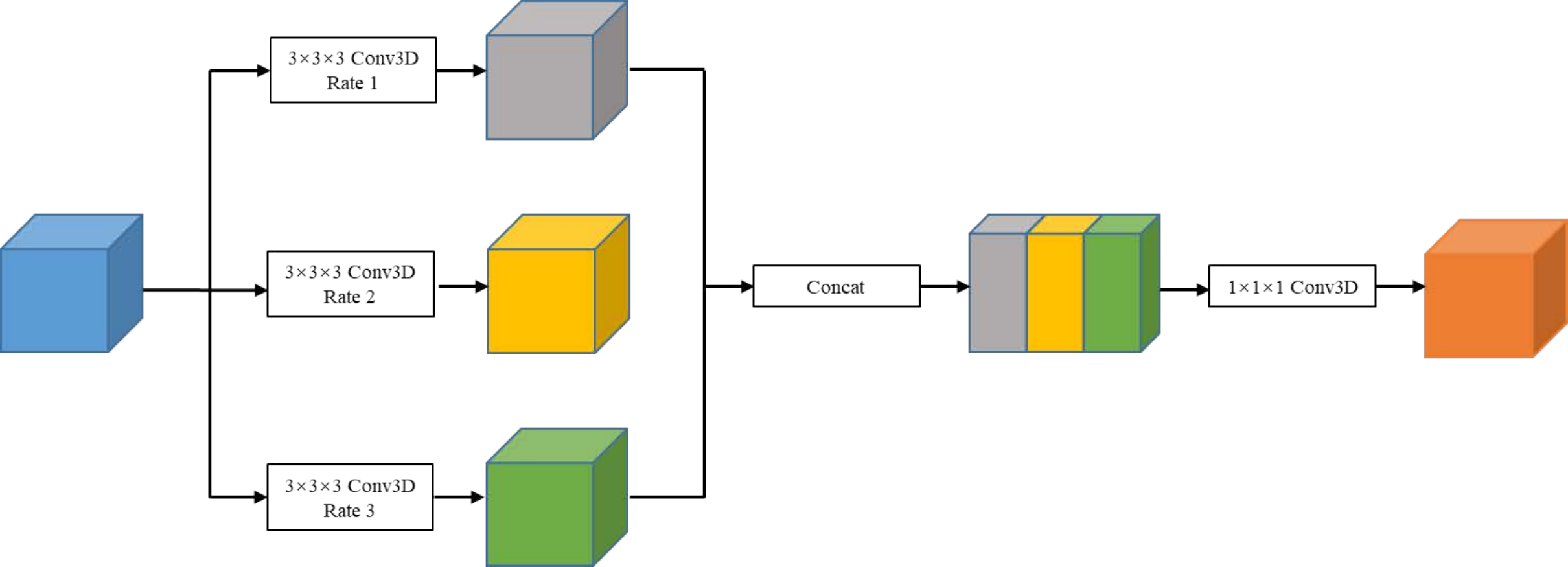}
	\caption{Illustration of the dilated convolution module.}
	\label{Fig:5}
\end{figure}

The structure of the 3D dilated convolution module is shown in Figure \ref{Fig:5}, which is formed by 3 parallel convolution branches. Then, the output feature maps of the different branches are concatenated together. Finally, $1 \times 1 \times 1$ convolution is performed to reduce the number of cuboid feature maps. The size of convolutional kernels of the three convolution branches is $3 \times 3 \times 3$, and the dilation rates are 1, 2, and 3, respectively. The size of the output feature maps in each branch keeps unchanged through the padding mechanism. For a $48 \times 48 \times 48$ voxels input cuboid, the LSC voxels account for about $\frac{1}{200}$ of the input data. Therefore, the expansion rate in the 3D dilated convolutional layer should not be too large. After each convolution operation, batch normalization and rectified linear unit are used.

\paragraph{(2)} Multi-scale and multi-mode pooling module\par

\begin{figure}
	\centering
		\includegraphics[scale=.55]{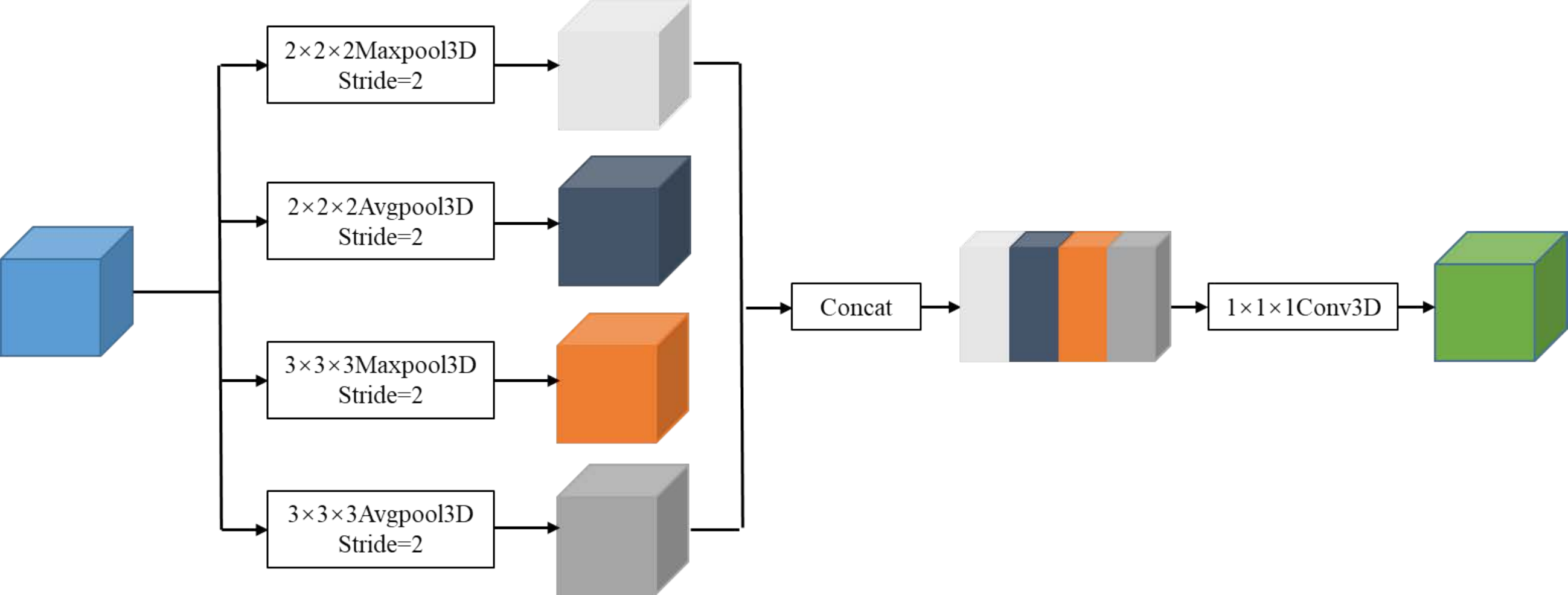}
	\caption{Illustration of the multi-scale and multi-mode pooling module. The size of the pooled feature map obtained by each branch is the same through the padding mechanism.}
	\label{Fig:6}
\end{figure}

The pooling operation downsample the data to reduce the computational cost of subsequent operations. While, it will also lead to the loss of spatial details. To relieve the information loss caused by the pooling operation, we propose a multi pooling module, which fuses the feature maps from pooling with different scales and models. The architecture of the multi pooling module is shown in Figure \ref{Fig:6}.\par

In the multi pooling module, we employ the maximum pooling, the average pooling of different sizes, the non-overlapping pooling \cite{krizhevsky2012imagenet}, and overlapping pooling modes. The average pooling can retain the background information of the image effectively, while, the maximum pooling can retain more texture information. The overlapping pooling causes overlap between adjacent blocks to extend, which will enhance the relationship between adjacent blocks. As shown in Figure 6, the multi pooling module has 4 pooling branches in parallel, which are $2 \times 2 \times 2$ maximum pooling, $2 \times 2 \times 2$ average pooling, $3 \times 3 \times 3$ maximum pooling, and $3 \times 3 \times 3$ average pooling. The feature map after each pooling branch has the same size due to the same stride are selected. Then the feature maps obtained from the 4 branches are spliced together. A $1 \times 1 \times 1$ convolution operation is performed to reduce the number of channels of the feature map.

\subsubsection{Decoder}
The decoder is mainly composed of two transposed convolutions, which gradually recovery the feature map to the same size as the input image. Skipping connections are employed to recover the detailed information of the prediction segmentation result in the decoding stage. The dilated convolution module at the bottom of the encoding stage outputs a $12 \times 12 \times 12$ feature map. The first transposed convolution up-sample the feature map from $12 \times 12 \times 12$ to $24 \times 24 \times 24$. The features of the second densely connected block are spliced with it. The 3D convolution is performed on the stitched features, and the obtained features are subjected to the second transposed convolution operation, and the up-sampled features are $48 \times 48 \times 48$. Finally, output the segmentation results of the voxels in the cuboid.

\subsubsection{Loss function}

In the training stage, we use the 3D-DSD proposed 3D deep supervision mechanism to promote network convergence. The deep supervision mechanism is to calculate the Segmentation error in hidden layers as a part of the total loss. Segmentation error of the hidden layers as the auxiliary loss and the output error of the last layer are joint together to constraint the network.\par

The joint loss function includes Dice-Sørensen Coefficient (DSC) \cite{dice1945measures} loss and cross-entropy loss. The DSC loss function is shown in (\ref{con:1}).

\begin{equation}
L(G,P)=1 - \frac{2 \times \sum_{i}^{n} p_i g_i}{\sum_{i}^{n} p_i + \sum_{i}^{n} g_i}\label{con:1}
\end{equation}
where $n$ represents the number of voxels, and $p_i$ $\in$ [0, 1] and $g_i$ $\in$ [0, 1] represent the voxel category in the predicted and ground-truth cuboid, respectively. Meanwhile, a weight $W$ is introduced for the cross-entropy loss function, as shown in (\ref{con:2}).

\begin{equation}
    W = 1- \frac{N_k}{N_c}\label{con:2}
\end{equation}
where $N_k$ is the number of annotated voxels, and $N_c$ is the number of voxels in the sub-volume. The cross-entropy loss function is shown in (\ref{con:3}), where $n$ is the number of voxels.

\begin{equation}
    L(p,\hat{p_i}) = - \frac{1}{n} \sum_{i}^{n} W_i (p_i \log \hat{p_i})\label{con:3}
\end{equation}

The joint loss function is shown as (\ref{con:4}).

\begin{equation}
    L=L(G,P) + \sum_{i}^{n} \lambda_k L_k(G,P) + L(p, \hat{p_i}) + \sum_{k}^{m} \lambda_k L_k(p, \hat{p_i})\label{con:4}
\end{equation}
where $\lambda$ is the hyperparameter of the branch network loss function, and $m$ is the number of hidden layers of deep supervision.

\subsection{Geometric calibration}
\begin{figure}
	\centering
		\includegraphics[scale=.55]{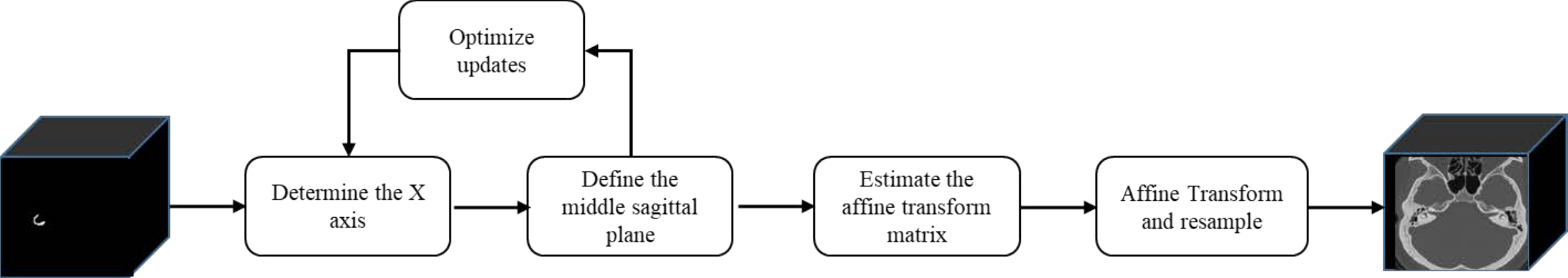}
	\caption{The flow chat of the calibration.}
	\label{Fig:7}
\end{figure}

Based on the segmentation of the lateral semicircular canal, we propose a calibration algorithm to eliminate the asymmetry of bilateral structures in the original temporal bone CT. The flow chart of the calibration is shown in Figure \ref{Fig:7}. In this calibration method, we define a 3D coordinate system. The three intersecting planes of the coordinate system are the median sagittal plane, the lateral semicircular canal plane, and the coronal plane. The median sagittal plane is the symmetry plane of the LSCs on both sides, the lateral semicircular canal plane can be fitted by the least square method, and the coronal plane can be calculated from the determined median sagittal plane and the LSCs plane.\par

Our calibration process is as follows:

\begin{enumerate}[(1)]
\item Determine the X-axis. According to the LSCs segmentation result, search the labeled voxels belong to the LSCs and find out the leftist point $P_1$ in the left LSC and the rightest point $P_2$ in the right LSC. The line crossing these two points is defined as the X-axis. The center point of the connecting segment of $P_1$ and $P_2$ is defined as the origin $P_0$.
\item Define the middle sagittal plane. The plane that orthogonal to the X-axis and across the origin $P_0$ is defined as the middle sagittal plane.
\item Optimize the update of $P_1$, $P_2$, and median sagittal plane of the LSCs. Ideally, the difference in vertical distance between the outermost points of LSCs and their corresponding middle sagittal plane is 0, but the actual human LSCs are not completely symmetrical, so the vertical distance difference is not queal to 0. We define an acceptable vertical distance difference $L_0$, and the vertical distance from the outermost point to the middle sagittal plane is defined as $L_1$. Firstly, initialize the outermost points $P_1$, $P_2$, if $L_1$ is less than $L_0$, then the $P_1$, $P_2$ are optimal, otherwise $P_1$, $P_2$ are reselected in LSCs, that is steps (1) and (2) are performed until $L_1$ is less than $L_0$.
\item Estimate the affine transform matrix. According to the normal vector of the median sagittal plane and the voxel matrix coordinate system of the original temporal bone CT, we estimate the affine transform matrix and transform the raw temporal bone in 3D to achieve automatic calibration of the original temporal bone CT. Specifically, the normal vector of the median sagittal plane is decomposed into three mutually perpendicular planes in the voxel matrix coordinate system, and the angle between the decomposition vector and the coordinate axis is calculated in each plane.
\item Affine transforms the CT slices and re-sample the calibrated temporal bone CT images by a predefined standard resolution. Such as 0.5mm to unify the voxel resolution in spatial and inter slices.
\end{enumerate}

\section{Experiments and discussions}

In this section, the results of the experiments are provided to show the performance of our proposed LSC segmentation algorithm and the effectiveness of the geometric calibration.\par

\subsection{Dataset}
With the approval of the Ethics Committee of Beijing Friendship Hospital affiliated to Capital Medical University, we collected 974 cases of raw temporal bone CT images. In which, 64 cases are manually calibrated and voxel-level labeled. All the private information on patients is encrypted. The size of the raw temporal bone CT image is $512 \times 512 \times N$ voxels ($N$ is about 180 \~{} 240 slices). The manual calibrated and labeled CT images are used to train and objective testing our segmentation network. The LSCs are annotated by clinically experienced doctors. In the 64 annotated cases, 52 cases are used as the training set, and the other 12 cases used as the testing set. During the training process, data argumentation is conducted by randomly rotating $-5^{\circ}$ \~{} $5^{\circ}$ in any direction. Meanwhile, it ensures the robustness of the training model to the segmentation of the LSC of the raw temporal bone CT. Then, the trained model is used to automatically segment the LSC in the raw temporal bone CT data, and the raw temporal bone CT is automatically calibrated using our proposed calibration method.

\subsection{Evaluation metric}

There are several evaluation metrics to measure the performance of medical image segmentation algorithms, in which Dice-Sørensen Coefficient(DSC) is a widely used metric index. We employ the dice coefficient to evaluate the segmentation performance of our method. The DSC is the similarity between the prediction $P$ in the segmentation task and the ground truth $G$.The definition of the DSC is shown in (\ref{con:5}). The larger the value of DSC, the higher the degree of overlap between the segmentation prediction and the ground truth, that is, the higher the segmentation accuracy.

\begin{equation}
    L(G,P) = \frac{2\mid P \cap G \mid}{\mid P \mid + \mid G \mid}\label{con:5}
\end{equation}

\subsection{Ablation experiments and analysis}
\begin{figure}[htbp]
\centering
\subfigure[3D-DCM]{
\begin{minipage}[t]{0.3\linewidth}
\centering
\includegraphics[scale=.75]{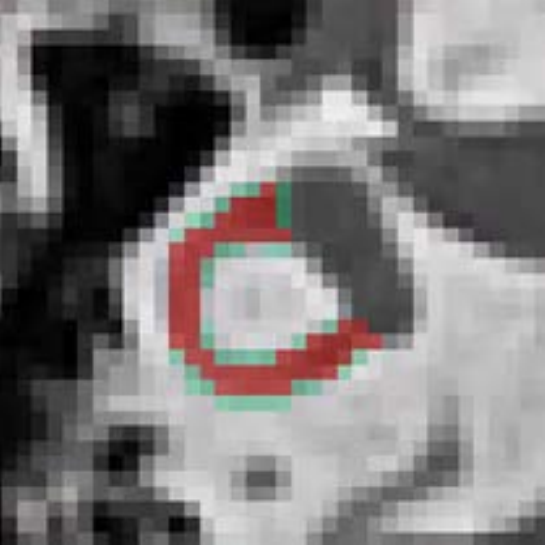}
\end{minipage}%
}%
\subfigure[3D-MP]{
\begin{minipage}[t]{0.3\linewidth}
\centering
\includegraphics[scale=.75]{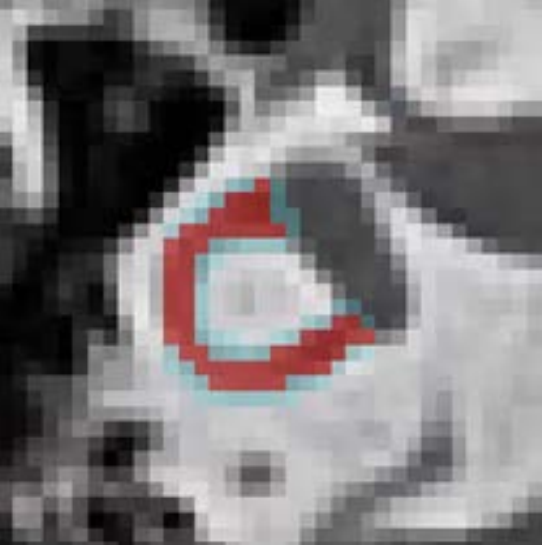}
\end{minipage}%
}%
\subfigure[3D-MFF]{
\begin{minipage}[t]{0.3\linewidth}
\centering
\includegraphics[scale=.75]{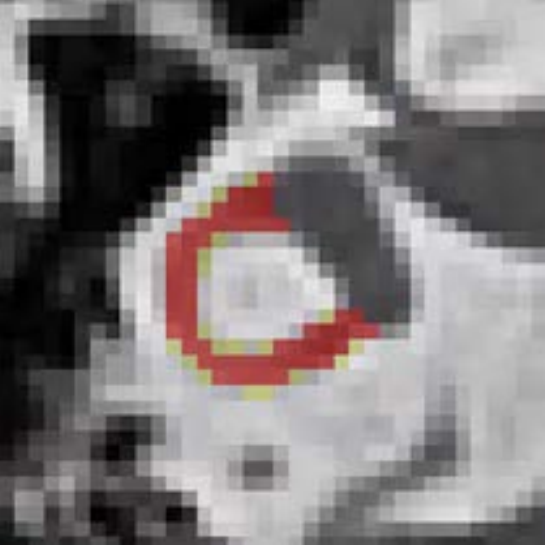}
\end{minipage}
}%
\centering
\caption{The subjective segmentation results of the 3D-DCM, the 3D-MP and the 3D-MFF. The red labels denote the ground-truth, the blue labels, blue labels and yellow labels denote the segmented results of 3D-DCM, 3D-MP and 3D-MFF respectively. We can see that (a) and (b) apt to over segmented and (c) achieved the test result.}
\label{Fig:8}
\end{figure}

To verify the effectiveness of our proposed method, we conduct ablation experiments to evaluate the 3D dilated convolution module and the multi pooling module. We designed 4 experiments in this section. Firstly, our previous work, 3D-DSD, was used as a benchmark. Secondly, we replace the last dense connection module (called 3D-DCM) in the encoding stage of 3D-DSD with our dilated convolution module. Thirdly, replace the pooling module in 3D-DSD with a multi pooling module(called 3D-MP). Finally, the dilated convolution module and the multi pooling module are simultaneously introduced in 3D-DSD, namely our proposed 3D-MFF. The segmentation performance of 4 schemes are shown in Table 1. And The visualization results of the ablation experiment are shown in Figure \ref{Fig:8}. In the segmentation of the LSCs, the segmentation results of the segmentation model often suffer from over-segmentation. The red labels denote the ground-truth, the blue labels, blue labels and yellow labels denote the segmented results of 3D-DCM, 3D-MP and 3D-MFF respectively. It can be seen from the visualization results of the ablation experiment that our proposed method has a good overlap with the ground-truth.

\begin{table}[h]
\centering
\begin{tabular}{l l l}
\hline
\textbf{Scenarios} & \textbf{DSC}\\
\hline
3D-DSD  & 70.03\%\\
3D-DCM  & 71.99\% \\
3D-MP   & 70.40\% \\
3D-MFF  & 72.23\% \\
\hline
\end{tabular}
\caption{The DSC performance of the ablation experiments.}
\end{table}

\subsubsection{Effectiveness analysis of dilated convolution module}

In the encoding stage, we retain the first and second dense connection modules in 3D-DSD, and the third densely connected module is replaced with a dilated convolutional module. Comparing the segmentation performance of the LSC of 3D-DSD and 3D-DCM, we observed that the DSC of 3D-DCM increased from 70.03\% of 3D-DSD to 71.99\%. This proves that the dilated convolution module can improve segmentation accuracy. We analyze the convolution operation with different expansion rates in the dilated convolution module, which effectively captures the multi-scale information of the target and which contributes to the improvement of segmentation performance.

\subsubsection{Effectiveness analysis of multi-scale and multi-mode pooling module}
Compared with the multi-mode pooling features fusion strategy in 3D-DSD, we use multi-scale and multi-mode pooling features fusion in the pooling module. From the experimental results, the segmentation of 3D-MP is improved by 0.37\% compared with 3D-DSD. Although the pooling operation can reduce the data dimension, it will also lose detailed information. The average pooling operation can retain the background information of the image, while the maximum pooling operation retains more texture information. Meanwhile, most of the pooling operations now use non-overlapping pooling. We use overlapping pooling to compensate for the split between image blocks. We combine average pooling and maximum pooling, while using non-overlapping pooling and overlapping pooling, so that the pooled features retain richer information to improve the accuracy of segmentation.

\subsection{Comparison with other networks}

\begin{figure}[htbp]
\centering
\subfigure[3D-Unet]{
\begin{minipage}[t]{0.3\linewidth}
\centering
\includegraphics[scale=.75]{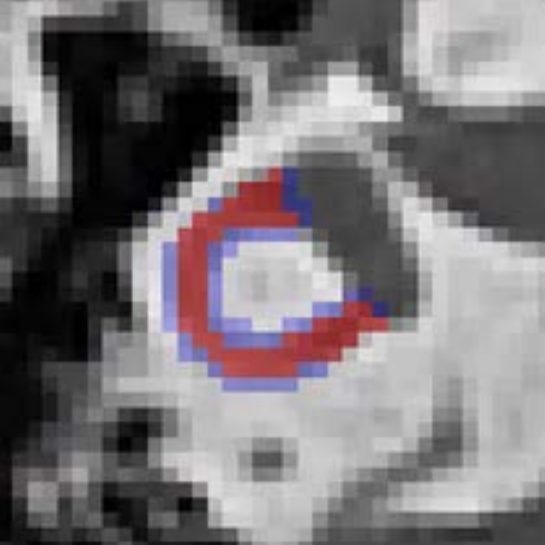}
\end{minipage}%
}%
\subfigure[3D-DSD]{
\begin{minipage}[t]{0.3\linewidth}
\centering
\includegraphics[scale=.75]{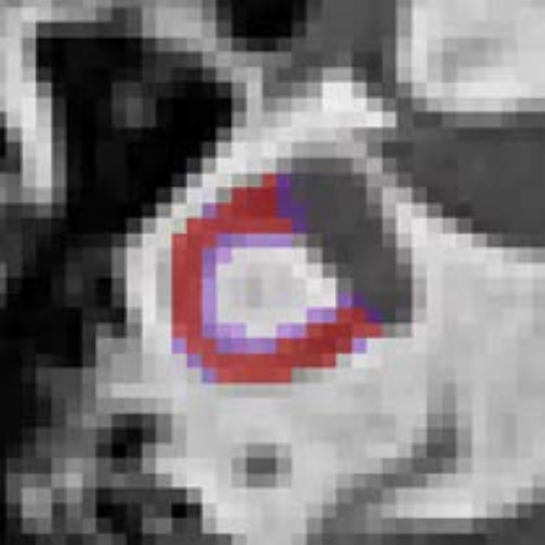}
\end{minipage}%
}%
\subfigure[3D-MFF]{
\begin{minipage}[t]{0.3\linewidth}
\centering
\includegraphics[scale=.75]{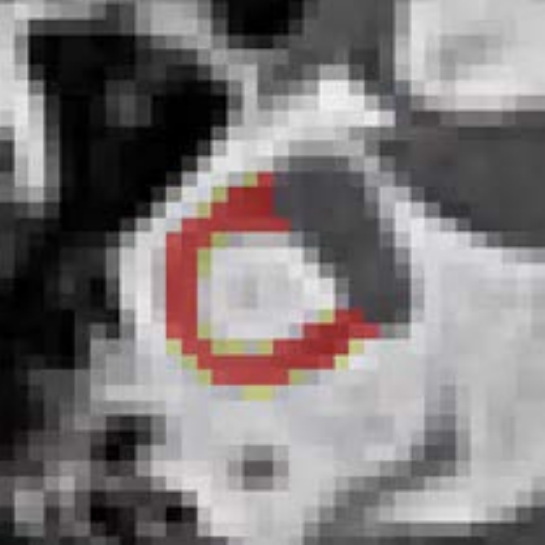}
\end{minipage}
}%
\centering
\caption{The subjective results of our proposed method, the 3D-Unet, and the 3D-DSD. The red labels denote the ground-truth, blue labels, purple labels and yellow labels denote the segmented results of 3D-Unet, 3D-DSD and 3D-MFF respectively. We can see that the result of our 3D-MFF is the best.}
\label{Fig:9}
\end{figure}

In this section, the proposed method is compared with 3D-Unet and 3D-DSD. The results of these two methods are shown in Table 2. The results show that the proposed method has higher segmentation accuracy than the other two methods. This improvement is attributed to our proposed 3D dilated convolutional module and the multi pooling module. The visualization results of our proposed method and other methods are shown in Figure \ref{Fig:9}. The subjective results of our proposed method, the 3D-Unet, and the 3D-DSD. The red labels denote the ground-truth, blue labels, purple labels and yellow labels denote the segmented results of 3D-Unet, 3D-DSD and 3D-MFF respectively. We can see that the result of our 3D-MFF achieved the best.

\begin{table}[h]
\centering
\begin{tabular}{l l l}
\hline
\textbf{Method} & \textbf{DSC}\\
\hline
3D-Unet  & 57.99\% \\
3D-DSD    & 70.03\% \\
3D-MFF   & 72.23\% \\
\hline
\end{tabular}
\caption{The DSC performance of the different segmentation methods.}
\end{table}

3D-Unet is a 3D version of the U-net, which is used for organ segmentation in 3D medical imaging. We employed the 3D-Unet to segmentation our testing LSC CT images, with an average DSC of 57.99\%.\par

Our previous work 3D-DSD is used to segment 9 important structures in the temporal bone. It is based on 3D-Unet, which uses dense connection blocks in the encoding stage, and uses a multi-mode pooling feature fusion strategy. The deep supervision mechanism assists in network training. The average DSC of 9 key structures is 77.18\%, but for the LSC, the DSC is 70.03\%. This may result from the more complicated structure of the LSC.

\subsection{Analysis of automatic calibration results}

\begin{figure}
	\centering
		\includegraphics[scale=.75]{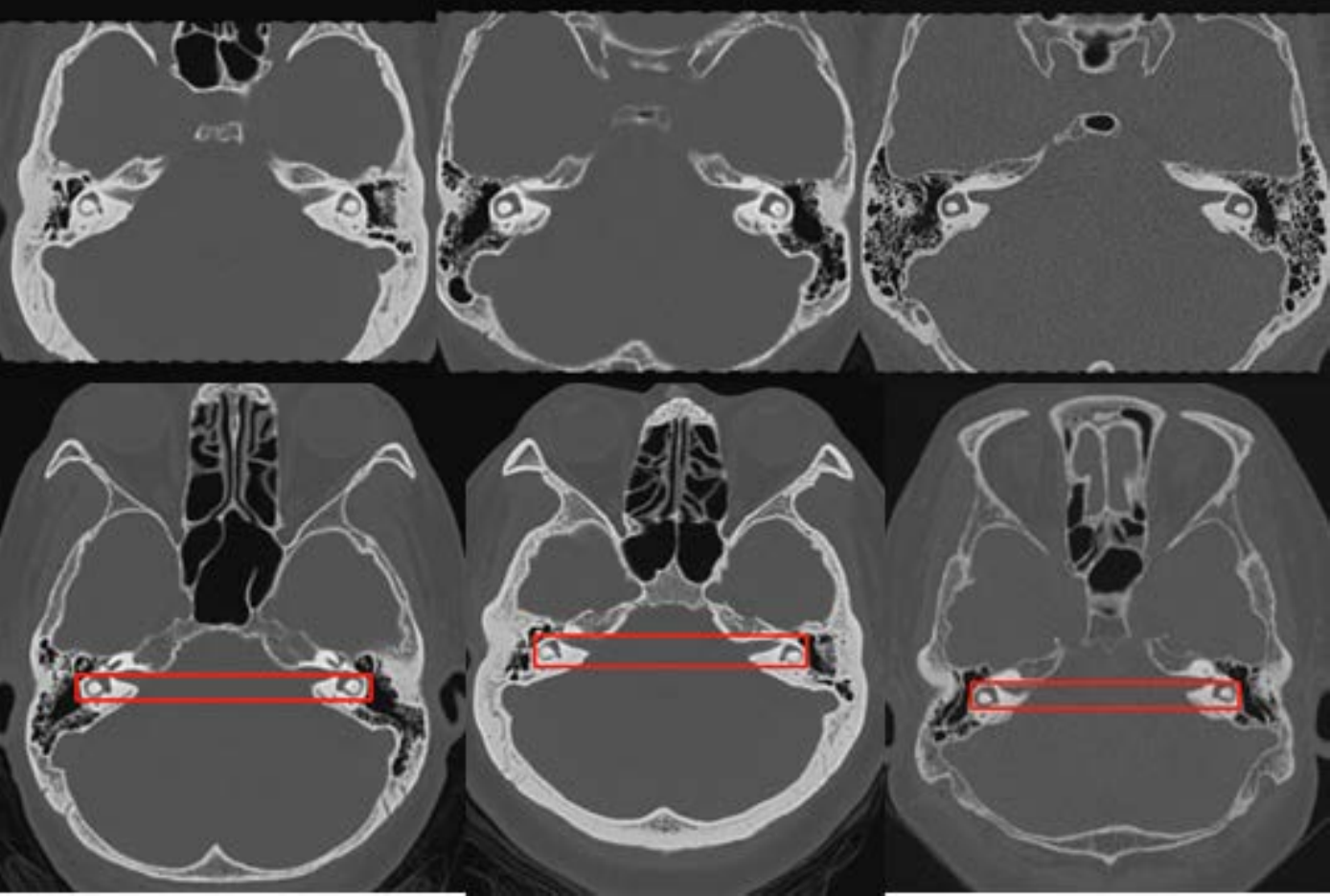}
	\caption{Comparison on the manually calibration (the upper row) and the automatic calibration (the bottom row).}
	\label{Fig:10}
\end{figure}

We use the trained 3D-MFF segmentation model to automatically segment the lateral semicircular canal in the 64 raw temporal bone CT, and then use an automatic calibration algorithm to calibrate these 64 cases of raw temporal bone CT images, of which the subjective effect is good by 45 cases, 19 cases with poor, the subjective calibration rate reaches 70.31\%. As shown in Figure \ref{Fig:10}, the images in the first row are manual calibrated results, the second row are the calibrated by our automatic calibration method.\par

To further evaluate the generalization of our proposed geometric calibration method, we test our method with 910 unlabeled raw temporal bone CT images. We evaluate the result by review the cross-sectional plane to check if the left and right LSCs appear on the same axial slice symmetrically. We defined the subjective evaluation results as three ranks, namely Excellent, Good, and Failed. The descriptions of these 3 ranks as follows.\par

Excellent: The left and right LSCs appear in the same axial slice, and perfectly symmetrical.\par

Good: The left and right LSCs appear in the same axial slice, but not perfectly symmetrical.\par

Failed: The left and right LSCs have not appeared in the same axial slice.\par

The statistical results are shown in Table 3. We can see that our proposed geometric calibration method for human temporal bone CT images achieved 777 acceptable results in 910 cases. The acceptable results that ranked excellent and good are 85.43$\%$.

\begin{table}[h]
\centering
\begin{tabular}{l l l}
\hline
\textbf{Rank} & \textbf{Ratio}\\
\hline
Excellent & 29.98\% \\
Good      & 55.45\% \\
Failed    & 14.57\% \\
\hline
\end{tabular}
\caption{Statistical results of the geometric correction method for temporal bone CT images.}
\end{table}

\begin{figure}[htbp]
\centering
\subfigure[Excellent]{
\begin{minipage}[t]{0.3\linewidth}
\centering
\includegraphics[scale=.75]{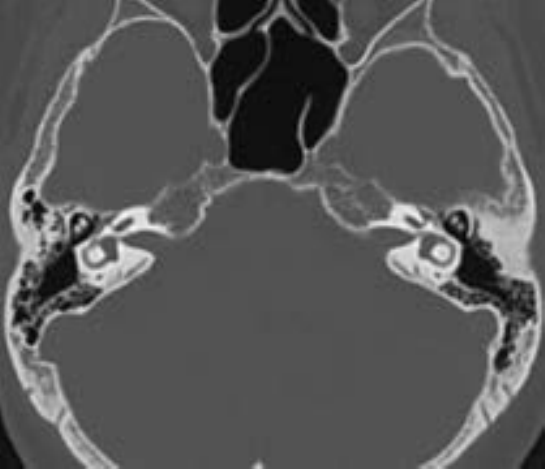}
\end{minipage}%
}%
\subfigure[Good]{
\begin{minipage}[t]{0.3\linewidth}
\centering
\includegraphics[scale=.75]{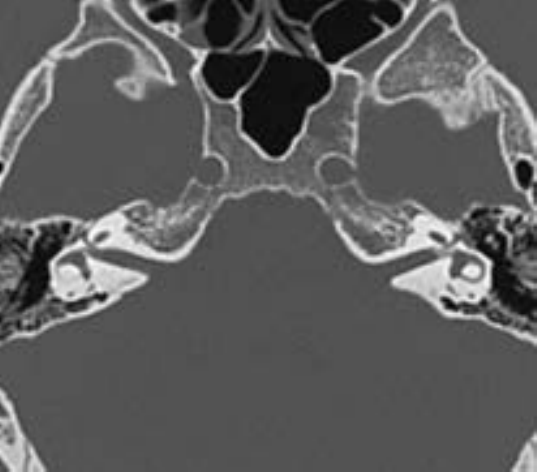}
\end{minipage}%
}%
\subfigure[Failed]{
\begin{minipage}[t]{0.3\linewidth}
\centering
\includegraphics[scale=.75]{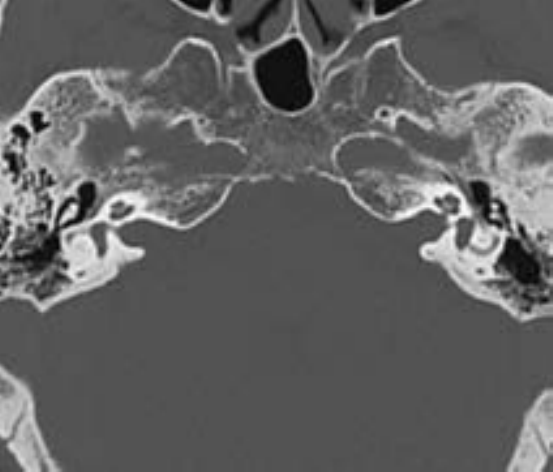}
\end{minipage}
}%
\centering
\caption{Three ranks of calibration for temporal bone CT images.}
\label{Fig:11}
\end{figure}

Some example results for 3 ranks are shown in Figure \ref{Fig:11}. We can see that In (a), the LSCs appear in a slice at the same time, and are bilaterally symmetrical; in (b), although the LSCs canals are bilaterally symmetrical, they do not appear in the slice at the same time, but differ by 1 \~{} 2 slices; although the LSCs in (c) is bilaterally symmetrical, the LSCs are not complete.\par

Note that we have not found similar research on the geometric calibration for temporal bone CT images. Therefore, we have not compared our geometric calibration method to other algorithms. We hope that more attention will pay for this topic in the future.

\section{Conclusion}

Aiming at the problem of bilateral asymmetry in the raw temporal bone CT images, we propose an automatic geometric calibration method for temporal bone CT images. In which, the LSC segmentation is vitally important in the calibration process. To improve the segmentation accuracy of the LSC, we designed a novel LSC segmentation network, which we introduce 3D dilated convolution module and a novel multi pooling module. The 3D dilated convolution module can extract multi-scale features efficiently. The multi pooling module can fusion richer details in different scales and modes to compensate for the loss information via pooling. To geometric calibrate the temporal bone CT images, a standard Cartesian coordinate system is defined and a method to align raw CT images with the standard coordinate system is proposed. Experimental results show the effectiveness of our proposed LSC segmentation algorithm and the geometric calibration method.\par

The work addressed in this paper has great potential to reduce the workload of radiologists and provides a fundamental milestone for computer-aided temporal bone CT analysis.

\section*{Acknowledgment}
This work is supported by the National Natural Science Foundation of China (grant number 61527807 and 61871006), the Science and Technology Development Program of Beijing Education Committee (grant number KZ201810005002).

\bibliographystyle{elsarticle-num}
\bibliography{sample.bib}







\end{document}